\documentclass[a4paper,10pt]{article}

\usepackage[english]{babel}
\usepackage{graphicx}
\usepackage[colorlinks, linkcolor=black, citecolor=black, urlcolor=black]{hyperref}
\usepackage{geometry}
\usepackage{todonotes}
\usepackage{ifthen}
\usepackage[frozencache=true, cachedir=minted-cache]{minted}
\usepackage{amsmath}
\usepackage{marvosym}
\usepackage{amssymb}
\usepackage{stmaryrd}
\usepackage{iris}
\usepackage{csquotes}
\usepackage{material/material}
\usepackage{makecell}

\usepackage{biblatex}
\newcommand{\ie}{i.e.}
\newcommand{\eg}{e.g.}

\begin{document}
\begin{titlepage}
    \newpage
    \thispagestyle{empty}
    \frenchspacing
    \hspace{-0.2cm}
    \includegraphics[height=3.4cm]{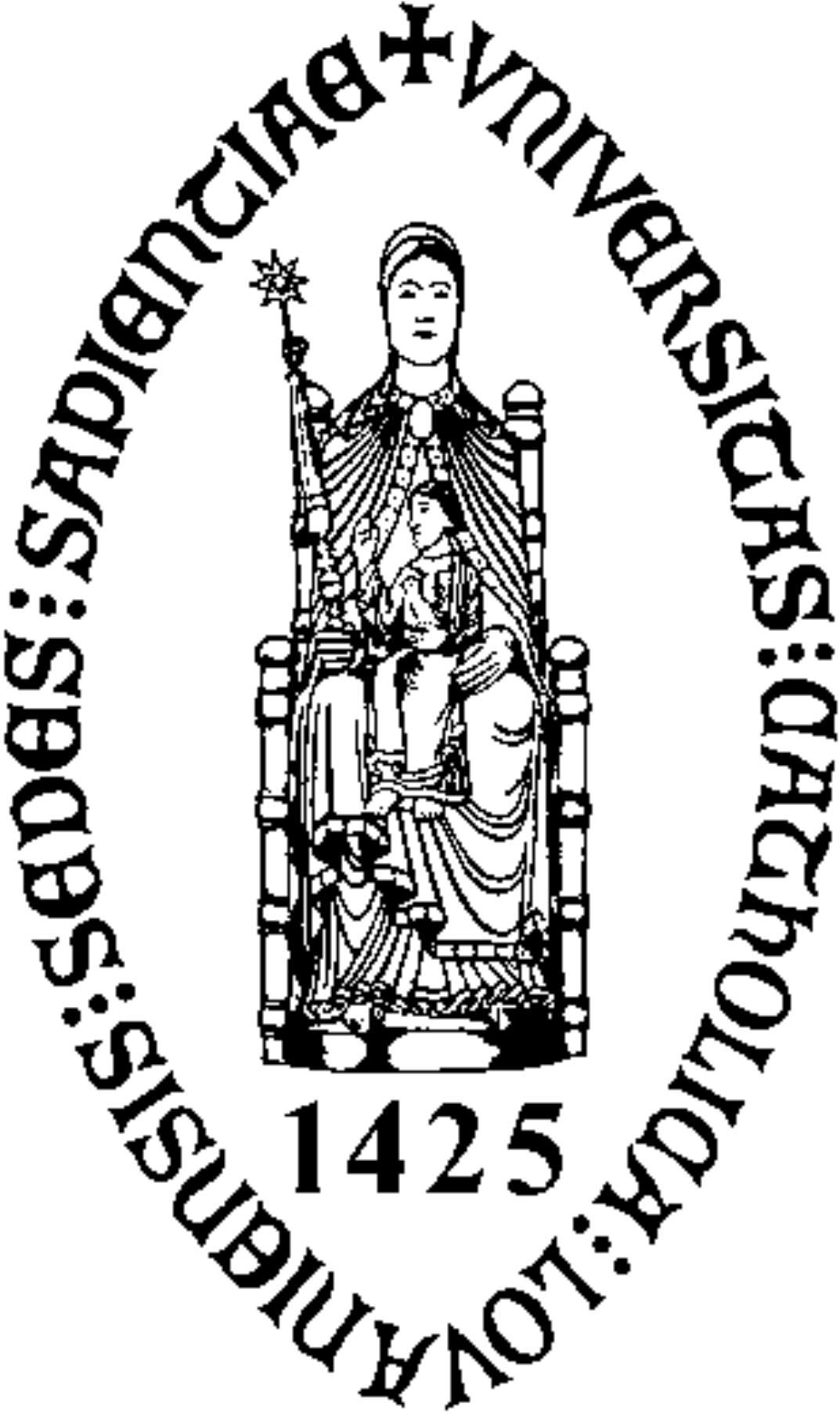}
    \hspace{0.2cm}
    \rule{0.5pt}{3.4cm}
    \hspace{0.2cm}
    \begin{minipage}[b]{8cm}
        \Large{Katholieke\newline Universiteit\newline Leuven}\smallskip\newline
        \large{}\smallskip\newline
        \textbf{Department of\newline Computer Science}\smallskip
    \end{minipage}
    \hspace{\stretch{1}}
    \vspace*{3.2cm}\vfill
    \begin{center}
        \begin{minipage}[t]{\textwidth}
            \begin{center}
                \LARGE{\rm{\textbf{\uppercase{Modular Formal Verification of Rust Programs with Unsafe Blocks}}}}\\
                \Large{\rm{Technical Report}}
            \end{center}
        \end{minipage}
    \end{center}
    \vfill
    \hfill\makebox[8.5cm][l]{%
        \vbox to 7cm{\vfill\noindent
            {\rm \textbf{Nima Rahimi Foroushaani}}\\
            {\rm \textbf{Bart Jacobs}}\\
            {\rm imec-DistriNet Research Group, KU Leuven, Belgium}\\
            {\rm \{nima.rahimiforoushaani, bart.jacobs\}@kuleuven.be}\\
            [2mm]
            {\rm Dec. 2022}
        }
    }
\end{titlepage}

\begin{abstract}
\emph{\rust{}} is a modern systems programming language whose type system guarantees memory safety. For the sake of expressivity and performance it allows programmers to relax typing rules temporarily, using \kwunsafe{} code blocks. However, in \kwunsafe{} blocks, the burden of making sure that the code does not end up having undefined behaviour is on the programmer. Even most expert programmers make mistakes and a memory safety bug in an \kwunsafe{} block renders all the type system guarantees void. To address this problem we are trying to verify soundness of \rust{} \kwunsafe{} code applying our \emph{Modular Symbolic Execution} algorithm. This text outlines our approach and the progress that has been made so far.
\end{abstract}

\tableofcontents
\newpage

\section{Introduction}
\rust{} is a relatively new programming language that provides memory safety without sacrificing performance and practicality, all the while being suited for systems programming as well. To achieve these all together has turned out not to be that easy, however. Other common programming languages usually trade these goals, one to another. Fortunately, \rust{}'s safety claims have been proven to be legitimate. The formal works, \emph{\rustBelt{}} \cite{Jung2017rustbelt}, \emph{\rustHorn{}} \cite{Matsushita2020rusthorn}, and \emph{\oxide{}} \cite{Weiss2019oxide} have proven the safety of formal languages, designed to capture the central characteristics of \rust{}. At the same time, \rust{} has proven it is not just a research language. It has found its way to the wild and in fact, is getting ever more popular. It shows the language is simple enough for developers and industry to use. So, it is completely fair to say \rust{} shows great promises. The main reason behind this success is the language type system. \rust{}'s type system leverages \emph{ownership} and \emph{borrowing} to rule out the possibility of simultaneous \emph{mutation} and \emph{aliasing}. In this way it prevents many common mistakes, developers commit regarding memory safety. The type system also makes \rust{} needless of a garbage collector which makes it suitable for embedded systems and systems programming.

It seems everything about \rust{} is perfect and it addresses all of the problems. But, does it? In the end, mutation and aliasing together are crucial whenever communication between threads is required, \eg{} \rustinl{Mutex}es. Programs that do reference counting, also need simultaneous mutation and aliasing. To provide a complete set of expected functionalities from a modern programming language and performance improvements, \rust{} introduces \kwunsafe{} code blocks. The type checker checks these blocks with some relaxations to allow the implementation of such functionalities. The cost of these relaxations is that programmers themselves should make sure the \kwunsafe{} blocks do not cause the program to exhibit \emph{undefined behaviour} (\ubehav{}). Developers abstract \kwunsafe{} blocks behind \emph{safe abstraction}s to prevent them from propagating through the codebase and to make them easier to inspect and reuse. It is effective but not enough. There have been memory safety bugs found in many \rust{} libraries \cite{Bae2021rudra}, including even the standard library \cite{Jung2017mutexguard_unsound} indicating keeping \kwunsafe{} blocks safe, is not that straightforward.

Before going further, some terminology agreements are necessary. In this text, \kwunsafe{} code refers to code enclosed in an \kwunsafe{} block. It does \textbf{not} mean there is necessarily something wrong with the code and does \textbf{not} mean the code's behaviour is necessarily undefined. Following \emph{\name{The Rustonomicon}} \cite{rustonomicon}, to refer to \kwunsafe{} code that shows \ubehav{}, we use \emph{unsound} \kwunsafe{} code in contrast to \emph{sound} \kwunsafe{} code which we know for sure would not exhibit \ubehav{}.

To address the problem of unsound \kwunsafe{} code in \rust{}'s ecosystem we plan to verify the safety of \rust{} programs with \kwunsafe{} code using \emph{\verifast{}}'s \cite{verifast} \emph{Modular Symbolic Execution} (\msexec{}) algorithm. \verifast{} is a research tool for verifying the safety and functional properties of \C{} and \java{} programs using \msexec{}. We apply the \msexec{} algorithm with the assumption that the input code has already passed \rust{}'s type and borrow checker. The outcome of this verification algorithm for a program would be finding potential problems or to guarantee that despite having \kwunsafe{} code, the program does not exhibit \ubehav{}. That is, no execution of the program accesses unallocated memory or contains data races. We represent and use the information needed for reasoning about program safety during the \msexec{} in the form of formulas of a dialect of \sepLogic{}. \sepLogic{} is a logic developed specifically for reasoning about pointer-manipulating computer programs. We get this required information from two main sources.
\begin{enumerate}
    \item we translate \rust{}'s rich type system's meaning into \sepLogic{} formulae. For the meaning of \rust{}'s types we are relying on the semantics provided by \rustBelt{}. Encoding \rustBelt{}'s semantics to make it usable by \verifast{} for verifying \rust{} programs is the novel aspect of this envisaged approach.
    \item we use the user-provided information in form of \sepLogic{} formulae annotated in the program code. The user can also guide the \msexec{} algorithm with lemmas and ghost commands to verify more programs.
\end{enumerate}
To evaluate our approach, we are extending \verifast{} to support \rust{} programs. We use \verifast{}'s backend as the underlying engine for \msexec{} and reasoning about \sepLogic{} formulae. It is worth noting, \verifast{} uses its own dialect of \sepLogic{}.

In the rest of this text, in Section \ref{sec:rust_unsafe} we take a tour of \kwunsafe{} \rust{} code, safe abstractions, and their potential unsoundness. Next, in Section \ref{sec:mod_sym_exec} we explain the \msexec{} algorithm for verifying the soundness of \rust{} programs with \kwunsafe{} blocks. In Section \ref{sec:rustbelt} we give a brief introduction to \rustBelt{}, its semantic model, and its approach to proving the soundness of safe abstractions. We also talk about the reasons we chose to use \rustBelt{}'s semantics and we show how we are going to use its semantic model in the \msexec{} algorithm. Next, in Section \ref{sec:impl} we report the progress that has been made so far to implement the suggested \msexec{} algorithm and we discuss why our approach provides added values with respect to \rustBelt{}. In Section \ref{sec:future} we explain the possibilities we envisage to contribute more to the safety of \rust{} ecosystem in the future. Eventually, we conclude in Section \ref{sec:conc}.

\section{Unsafe Code and Safe Abstractions}\label{sec:rust_unsafe}
To guarantee memory safety \rust{} types represent \emph{ownership}. Listing \ref{listing:rust_access_type} shows these different types of ownership of a vector. The most basic type of ownership is \emph{exclusive ownership}. Owner variables, \eg{} \rustinl{v}, have exclusive ownership. An active owner variable rules out aliasing entirely. The value is represented in the whole program just by its owner and gets dropped when the owner is out of scope. We can transfer the ownership to other functions/threads. But it is still not that expressive. To regain ownership after passing it to a function, we should return it back which is very inconvenient in most cases. To solve this issue, instead of moving exclusive ownership we can \emph{borrow} it \emph{temporarily}. A mutable reference grants \emph{temporary exclusive ownership}. In our example, \rustinl{mrv} gives us write access. We mutate the vector inside function \rustinl{push_four} through the passed mutable reference, \rustinl{mrv}. Once \rustinl{mrv} is out of scope, the owner \rustinl{v} gets its exclusive ownership back again. Owners and mutable references, representing exclusive ownership, rule out aliasing. However, aliasing is needed to give access to multiple threads to the same memory location. To represent a part of memory and sub-parts of it at the same time is also very common and handy in programming. Shared references are the \rust{}'s answer for aliasing. Notice that we have a shared reference \rustinl{srv} to vector \rustinl{v} and a shared reference \rustinl{first} to its first element at the same time. To preserve memory safety shared references rule out mutation.

All of the references in \rust{} have a \emph{lifetime} in their type. \rustinl{'l} in the type \rustinl{&'l mut i32} is a lifetime. Lifetimes represent a continuous range of program execution steps. Type system's guarantees about references hold, as long as their lifetime is alive. Look at the signature of the function \rustinl{push_four}. It has a lifetime parameter \rustinl{<'a>} which is used in the type of parameter \rustinl{r}, \ie{} \rustinl{&'a mut Vec<i32>}. Lifetime parameters are the way callees get informed about the aliveness of a lifetime in the caller. They are \textcquote{thebook}{another kind of generics}, in the sense that they are not run-time variables. They get instantiated at compile-time, \ie{} when we call a function with a lifetime parameter, the compiler tries to find a suitable lifetime instantiation for the lifetime parameter. In our example, the lifetime that \rustinl{mrv} has in its type, has been annotated using comments in the code, \rustinl{l1}. It is a suitable lifetime for instantiating \rustinl{push_four}'s lifetime parameter. One implicit type system's guarantee about lifetime parameters is that they all \emph{outlive} the function's body lifetime.
\begin{listing}
\begin{minted}{Rust}
pub fn push_four<'a>(r: &'a mut Vec<i32>) {
    r.push(4)
}
/*** [l1] means the lifetime l1 */
pub fn access_types() {
    let mut v: Vec<i32> = vec![1, 2, 3]; // v is the owner
    {//----------------------------------------------------
        let mrv: &mut Vec<i32> = &mut v; //               |
        /***                                              |
         * mrv is a mutable borrow of v                   |
         * as long as this borrow is alive it            [l1]
         * is not possible to access                      |
         * the vector through v                           |
         */ //                                            |
        push_four(mrv); // mutable borrow has full access |
    }//----------------------------------------------------

    let _ = v.pop(); // v has its ownership back

    {//----------------------------------------------------
        let srv: &Vec<i32> = &v; //                       |
        /***                                              |
         * srv is a shared/immutable borrow of v          |
         * the vector cannot get mutated as long as       |
         * it is borrowed by any immutable borrow         |
         */ //                                            |
        {//----------------------------------------       |
            let first: &i32 = //                  |       |
                v.first().unwrap(); //            |       |
            /***                                  |      [l2]
             * multiple shared references,        |       |
             * borrowing from the same owner,     |       |
             * can coexist                       [l3]     |
             */ //                                |       |
            println!("{} is the first in {:?}", //|       |
                first, srv); //                   |       |
        }//----------------------------------------       |
    }//----------------------------------------------------
    let _ = v.pop();
    /***
     * The owner v goes out of scope here
     * and the value gets dropped
     */
}
\end{minted}
\caption{Different types of memory ownership in \rust{}'s types}
\label{listing:rust_access_type}
\end{listing}

\rust{}'s type system rules out simultaneous mutation and aliasing using the ownership and borrowing rules. However, communication between threads needs mutation and aliasing together. As an example consider a \rustinl{Mutex}. We need to have references to it in different threads, aliasing, and we need to lock it in those threads, mutation. To have mutation and aliasing of a memory location in a program simultaneously is against \rust{}'s type system rules. Moreover, the safety checks to maintain the type system's guarantees are necessarily conservative and valid programs that do not pass these checks are not that few. To address expressivity besides safety \rust{} introduces \kwunsafe{} code, \ie{} code blocks annotated with the \kwunsafe{} keyword. The method \rustinl{set} in Listing \ref{listing:cell} is an example of using an \kwunsafe{} code block. \kwunsafe{} code still gets checked by the type and borrow checker, but with some relaxation. The \rustbook{} \cite{thebook} book mentions five actions you can take just in \kwunsafe{} code and calls them \emph{unsafe superpowers}. Three of these unsafe superpowers are inherently unsafe primitive constructs and two of them are just indicating there are some other \kwunsafe{} parts inside.

In this project, among primitive unsafe constructs, we will initially focus on supporting \kwunsafe{} code involving \emph{dereferencing raw pointers}. The two others are used relatively rarely. Raw pointers are similar to \C{} pointers. \rust{}'s borrow checker does not track them and they can be null or dangling. Their types are of the form \rustinl{*const T} or \rustinl{*mut T} for arbitrary pointee type \rustinl{T}.

Among the two non-primitive superpowers, we are interested in \emph{call an \kwunsafe{} function/method}. An \kwunsafe{} function or method's signature is annotated with \kwunsafe{} keyword, \eg{} \rustinl{unsafe fn function() {...}}. The keyword \kwunsafe{} in the function's signature intuitively means calling this function has requirements that the type system cannot check and it is up to the programmer to make sure they have been met. An \kwunsafe{} function's body is an \kwunsafe{} code block. Using \kwunsafe{} functions propagates the \kwunsafe{} code to the callers.

\subsection{Safe Abstractions}
If we used \kwunsafe{} superpowers to implement a functionality we can expose the unsafety to the user code by marking our functions as \kwunsafe{}. But it should stop at some point. Otherwise, the \kwunsafe{} code propagates all over the codebase and we would not get much benefit from \rust{}'s type system. It puts the burden of safety checks on the programmer's shoulders and is in contradiction with type safety. It is much better to abstract the \kwunsafe{} parts in a safe function. Such a function would be a \emph{safe abstraction}. Then it can be called in safe \rust{} and the type system checks whether the caller meets the requirements the function type represents. In case of safe functions without any \kwunsafe{} block in their body, the type system also checks that the function body complies with the function type. However, it is not the case for a safe abstraction. It is the programmer's job to ensure the function body satisfies what the function type announces to the safe world.
As an example, let us look at Listing \ref{listing:cell}. The method \rustinl{set} is a safe abstraction. Notice that its signature is safe and it gets an argument of type \rustinl{&'a self} that is a shared reference to an object of \rustinl{struct Cell}. While it has only a shared reference to the object, using an \kwunsafe{} block and dereferencing a raw pointer, it writes to the contents of the object. The code mutates the contents of memory through a shared reference! It is in contradiction with the core rules of the type system. Recall that one of the guarantees of a shared reference type is that no mutation would happen during the reference's lifetime. But this \rustinl{set} method is not a horrible mistake. The fact that there is a shared reference together with the type system's guarantees implies there is a valid chunk of memory containing a valid \rustinl{Cell} value. If we could make sure all aliases of a \rustinl{Cell} object are limited to just one thread there would not be a memory safety issue. There are other type checks regarding sending ownership and borrows to other threads. Because of those checks and the code line \rustinl{impl !Sync for Cell {}} in our example, the type system does not allow sending a shared reference of a \rustinl{Cell} object to another thread. Moreover, no public method in \rustinl{Cell} library leaks a reference to the internal state of a \rustinl{Cell} object. That prevents sending \emph{deep pointers} of the \rustinl{Cell} to other threads. These together means library \rustinl{Cell} holds the following property: All aliases of a \rustinl{Cell} object remain in the same thread. That would be our \rustinl{Cell} library \emph{invariant}. The usage of \kwunsafe{} code in \rustinl{Cell} library is sound and abstracts away the \kwunsafe{} block. The library adds the functionality of mutation through shared reference, but because of its invariant, it is still safe. Safe code can use \rustinl{Cell} objects without the necessity of taking care of memory safety. Our example is close to what the real \rustinl{std::cell::Cell} in the standard library is. Libraries that abstract away their unsafe superpower application from their user, usually guarantee memory safety by holding such invariants. Mutating an object's internal state through shared references, abstracted from the user code, is called \emph{interior mutability} and \rustinl{std::cell::Cell} is the most basic form of interior mutability in \rust{}.
\begin{listing}
\begin{minted}{Rust}
pub struct Cell {
    value: i32,
}

impl Cell {
    pub fn new(value: i32) -> Cell {
        Cell { value }
    }

    pub fn get<'a>(&'a self) -> i32 {
        self.value
    }

    pub fn set<'a>(&'a self, n: i32) {
        let value_mut_ptr = &self.value as *const i32 as *mut i32;
        unsafe {
            *value_mut_ptr = n;
        }
    }
}
impl !Sync for Cell {}
\end{minted}
\caption{A simplified version of \rustinl{std::cell::Cell}}
\label{listing:cell}
\end{listing}

\subsection{Unsound Unsafe}
Not all \kwunsafe{} usages are sound. It is easy to use an unsafe superpower and end up with undefined behaviour (\ubehav{}). Recall that raw pointers are \C{}-style pointers and dereferencing a null or dangling raw pointer is \ubehav{}. Even worse, a safe abstraction's body may not satisfy the guarantees the function signature describes. Listing \ref{listing:unsound_unsafe} shows examples for both cases. The function \rustinl{breaks_ty_sys} in this example does not access unallocated memory. However, it violates the type system guarantees that type checker always assume when it checks safe code. In such cases, the problem might show up in the execution of safe code. In general, writing sound \kwunsafe{} code is very difficult, especially in the presence of \rust{} language constructs such as higher-order functions, traits and panics that complicate the task of analyzing the possible behaviors of a piece of code.
\begin{listing}
\begin{minted}{Rust}
pub fn deref_null() {
    let ptr = 0x0usize as *mut i32;
    unsafe {
        *ptr = 42;
    }
}

pub fn breaks_ty_sys(rrx: &mut &mut i32) {
    let ptr = rrx as *mut &mut i32 as *mut *mut i32;
    unsafe {
        *ptr = 0x0usize as *mut i32;
    }
}
\end{minted}
\caption{Unsound \kwunsafe{} code examples}
\label{listing:unsound_unsafe}
\end{listing}

\section{Modular Symbolic Execution (MSE)}\label{sec:mod_sym_exec}
\rust{} has a rich type system that checks memory safety statically. But its soundness relies on the soundness of the libraries that apply unsafe superpowers. Programmers who develop these libraries, being human, make mistakes. A single memory safety bug in an \kwunsafe{} block encapsulated in a library that is used by a program renders all of the type system's guarantees void. Here is the point we are targeting to contribute to \rust{} safety.
To verify soundness of safe abstractions and \kwunsafe{} code behind them, we propose applying \emph{Modular Symbolic Execution} (\msexec{}) on \kwunsafe{} containing parts of programs and observing if all the memory accesses through raw pointers are safe and if safe abstractions are right about what they suggest to the safe world by their interface types. The latter is, checking if safe abstractions implement exactly what their signature/type means. Here, arises a more fundamental question. What do \rust{} types mean? We need to answer this question before we could check the bodies of safe abstractions against their type's meaning. Fortunately, we do not need to propose an answer from scratch. \rustBelt{} \cite{Jung2017rustbelt} already suggests formal semantics for \rust{}'s types. In this section, we give a brief example-driven explanation of the Modular Symbolic Execution (\msexec) of \rust{} programs. Later, in Section \ref{sec:rustbelt} we briefly discuss \rustBelt{} \cite{Jung2017rustbelt}, a well-respected work that suggests a formal semantic model for \rust{}'s types. Moreover, we will explain why we have chosen to use its semantic model and we show a more sophisticated motivating example of the \msexec{} algorithm leveraging \rustBelt{}'s semantic model.

Listing \ref{listing:deque} shows parts of a library that implements a \emph{\name{Deque}} (double-ended queue) all using \kwunsafe{} code. This library's functions receive and return \name{Deque} instances just using raw pointers. In \rust{}, having a raw pointer does not guarantee anything about the memory it points to, \eg{} the type checker does not count on anything about the pointee of the returned raw pointer from \rustinl{create_deque}. That means trying to verify this example we would need to check \rustinl{create_deque}'s body against fewer type-induced proof obligations which simplifies the introduction to our \msexec{}. Later in \ref{ssec:rustbelt_and_mse}, we will discuss an example of \msexec{} of a safe abstraction, with types that represent more guarantees.
\begin{listing}
\begin{minted}{Rust}
use std::ptr::addr_of_mut;

pub struct Node {
    prev: *mut Node,
    value: i32,
    next: *mut Node,
}

pub unsafe fn create_deque() -> *mut Node {
    let sentinel: *mut Node = std::alloc::alloc(std::alloc::Layout::new::<Node>()) as *mut Node;
    if sentinel.is_null() {
        std::alloc::handle_alloc_error(std::alloc::Layout::new::<Node>())
    }
    addr_of_mut!((*sentinel).prev).write(sentinel);
    addr_of_mut!((*sentinel).next).write(sentinel);
    return sentinel;
}
// ...
\end{minted}
\caption{A \name{Deque}, implemented just using \kwunsafe{} \rust{}}
\label{listing:deque}
\end{listing}

\subsection{Concrete Execution}
We are trying to show no execution of \kwunsafe{} code performs memory access violations and neither violates the type system's guarantees. In the \name{Deque} example, it just suffices to make sure our implementation does not perform memory access violation. Let us assume we chose the most naive solution. We decide to verify the \name{Deque} by executing all of its possible executions and observe if they access memory chunks that they do not have any right to.

We execute our program on an abstract machine. \emph{\store{}} and \emph{\heap{}} together are the state of the machine. \store{} is a function that maps variables to their current value. \heap{} is an accounting of the abstract machine's memory. Mathematically, \heap{} is a \emph{multiset} of heap chunks. Heap chunks are predicates applied to arguments that represent information about the memory. We use predicates from \verifast{}'s dialect of \sepLogic{}. \sepLogic{} is a logic family, developed specifically for reasoning about pointer-manipulating concurrent programs. We will talk more about \verifast{} in Section \ref{sec:impl}.

Let us start by executing the \rustinl{create_deque} function. \store{} and \heap{} are empty at the beginning and the first statement is \rustinl{let sentinel: *mut Node = std::alloc::alloc(...) as *mut Node;}. From the documentation of \rustinl{std::alloc::alloc}, we know that if the function returns, either it has failed to allocate the requested memory and the return value is a \rustinl{null} raw pointer or it has allocated required memory in which case we know the following.
\begin{enumerate}
    \item The address stored in \rustinl{sentinel} is not \rustinl{null}
    \item The address stored in \rustinl{sentinel} is aligned
    \item Adequate number of bytes to store an instance of \rustinl{Node} are allocated at the address stored in \rustinl{sentinel}
    \item Up until deallocating this memory block, no other part of the program can allocate any of these bytes
\end{enumerate}
After the execution of this line, there are different possible machine states. In one state, the value in the \rustinl{sentinel} could be \rustinl{null}, in another one \rustinl{0x1000}, and in another one \rustinl{0x12345}. In the states where the \rustinl{sentinel}'s value is not \rustinl{null}, there are chunks, batches of bytes, allocated in \heap{} that our program is allowed to access. But since the memory has just been allocated, we do not know anything about the values stored in those bytes. The memory is not yet initialized after allocation and we do not have any guarantees about the validity of values stored in it. That is why we are representing them with the special value $\poison$. In \rust{} \emph{producing} an invalid value is considered \ubehav{}. \textcquote{rustonomicon}{Producing a value happens any time a value is assigned to or read from a place, passed to a function/primitive operation or returned from a function/primitive operation}. \textcquote{rustonomicon}{An integer [\dots], floating point value [\dots], or raw pointer obtained from uninitialized memory, or uninitialized memory in a \rustinl{str}} are invalid values. To reflect this, if a program attempts to read a $\poison$ value our execution algorithm gets stuck, \ie{} does not verify the program.

\begin{figure}
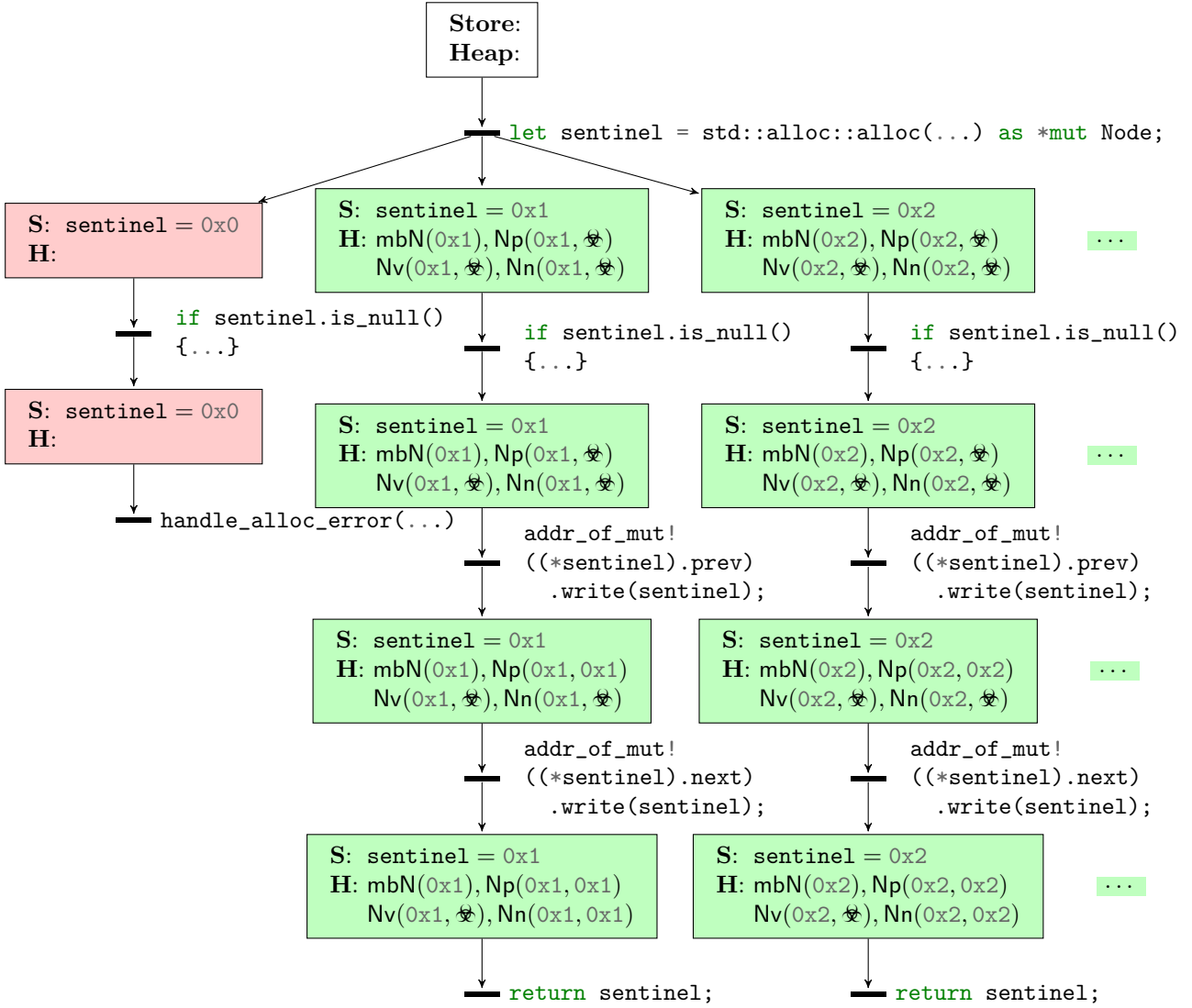

    \centering
    \drawConcExecTree{}
    \caption{The concrete execution tree of function \rustinl{create_deque} in Listing \ref{listing:deque}. The predicate names have been abbreviated in this figure as follows. $\mallocBlock{Node}\rightarrow\mbNode$, $\predName{Node\_prev}\rightarrow\nodePrev$, $\predName{Node\_value}\rightarrow\nodeVal$, and $\predName{Node\_next}\rightarrow\nodeNext$}
    \label{fig:conc_exec_tree}
\end{figure}

It is worth noting we do not want to verify our program against a specific concrete machine, and it means the set of possible addresses is practically infinite. Thanks to the non-determinism of the address that \rustinl{std::alloc::alloc(...)} returns, there are practically infinitely many possible states after executing this line of code. We can show program execution paths in a tree which branches whenever there are different possible outcome states after executing a statement. Figure \ref{fig:conc_exec_tree} shows the \emph{concrete execution tree} for \rustinl{create_deque}. We represent the information we know about the allocated block of memory in \heap{} using the following heap chunks.
\begin{enumerate}
    \item $\mallocBlock{Node}(\rustinmath{0x1})$ means there is an allocated block of memory starting from address \rustinl{0x1} with sufficient bytes to store an instance of \rustinl{Node}.
    \item $\predName{Node\_prev}(\rustinmath{0x1},\poison{})$ means the address \rustinl{0x1} plus offset of field \rustinl{prev} of \rustinl{struct Node} is an aligned memory address and points to enough bytes allocated to hold a value of the type of the field \rustinl{prev}, \ie{} \rustinl{*mut Node} and no other thread knows about this bunch of bytes, \ie{} we have write and read access to those bytes. The second argument, $\poison{}$, is the current value stored in those allocated bytes.
    \item $\predName{Node\_value}$ and $\predName{Node\_next}$ similar to $\predName{Node\_prev}$
\end{enumerate}

Looking at Figure \ref{fig:conc_exec_tree} we have an execution path in which \rustinl{sentinel==0x0}, marked by red and infinitely many execution paths, marked by green, in which \rustinl{sentinel!=0x0}, \ie{} the ones where memory allocation succeeded. In case of memory allocation failure, the program aborts by a call to \rustinl{std::alloc::handle_alloc_error(...)}. In case of successful allocation with the state with \rustinl{sentinel==0x1}, we have to execute the subsequent write operations.

\rustinl{addr_of_mut!((*sentinel).prev).write(sentinel);} is a write to field \rustinl{prev} of a \rustinl{Node} memory block at the address stored in \rustinl{sentinel}, on this path \rustinl{0x1}. This write is safe because in our \heap{} we have the predicate $\predName{Node\_prev(\rustinmath{0x1},\poison{})}$. After the write the value stored in the field gets updated, $\predName{Node\_prev(\rustinmath{0x1},\rustinmath{0x1})}$. If there was no such chunk in \heap{}, our execution algorithm would get stuck, representing that the program is attempting to access memory, without being sure that it has the right to do so. The next write operation is safe similarly. The final statement is \rustinl{return sentinel;}. Representing the return procedure involves many details. Since our goal here is to explain modular symbolic execution, we don't discuss possible cases and keep ourselves focused on this example. Here, the value of the local \rustinl{sentinel} gets copied into the return place. Notice that we still have the memory chunks produced in the \heap{}. The execution finished successfully and this path is fine. Note that, since the execution tree is (practically) infinite, traversing it entirely according to the procedure described here is (practically) impossible in finite time.

\subsection{Symbolic Execution}
\begin{figure}
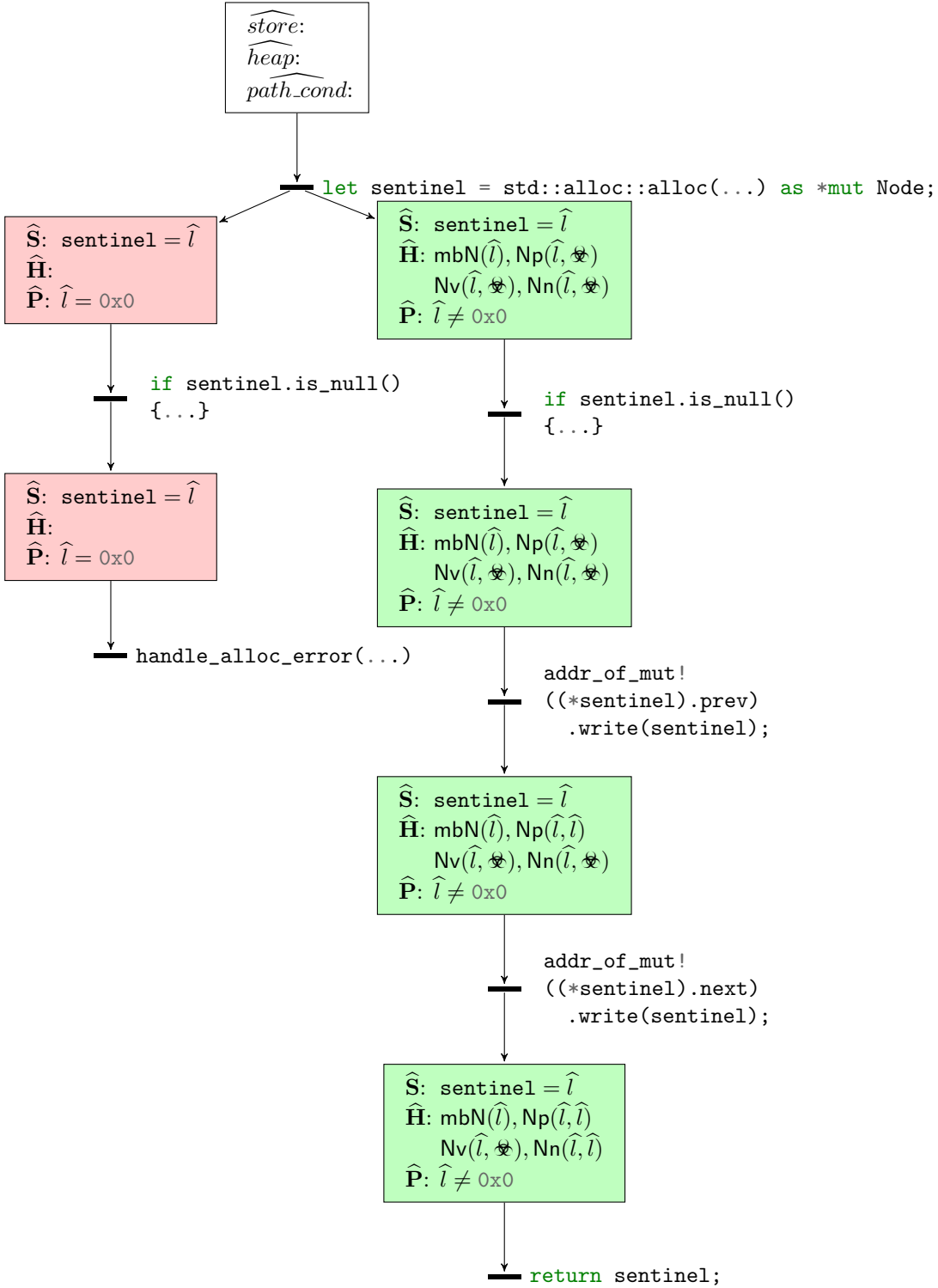

    \centering
    \drawSymExecTree{}
    \caption{The symbolic execution tree of function \rustinl{create_deque} in Listing \ref{listing:deque}. The execution paths represent the paths with the same colour in Figure \ref{fig:conc_exec_tree}. The predicate names have been abbreviated in this figure as follows. $\mallocBlock{Node}\rightarrow\mbNode$, $\predName{Node\_prev}\rightarrow\nodePrev$, $\predName{Node\_value}\rightarrow\nodeVal$, and $\predName{Node\_next}\rightarrow\nodeNext$}
    \label{fig:sym_exec_tree}
\end{figure}

Instead of dealing with infinite concrete execution trees, it is possible to abstract away some details that make paths distinct and represent infinitely many of them using a single one. To do so we use \emph{symbol}s instead of concrete values. Using symbols, we forget about corresponding concrete values, but we still remember the facts that hold for all of them. In this text, we typeset symbols like $\sym{sym}$, to make them distinct. Back to our example, to represent the address stored in \rustinl{sentinel} after allocation we choose a symbol, let us say $\sym{l}$, and also store the facts we know about it. We will have a single symbolic execution path for the case of allocation failure which in $\sym{l}=\rustinmath{0x0}$ and another symbolic execution path representing all the concrete paths where memory allocation is successful. In all of the successful paths, $\sym{l}\neq\rustinmath{0x0}$ and the \heap{} chunks at address $\sym{l}$ would be produced. To represent a symbolic execution state, we show the symbolic \store{} as $\sStore{}$, the symbolic \heap{} as $\sHeap{}$, and the \emph{path condition} as $\sEPK{}$. The path condition is our knowledge base about symbols. We store the persistent facts we know about symbols in it. Figure \ref{fig:sym_exec_tree} shows the finite \emph{symbolic execution tree} corresponding to the practically infinite concrete execution tree shown in Figure \ref{fig:conc_exec_tree}.

The execution using symbols and facts we know about them is called \emph{Symbolic Execution}. It is modelling of the concrete execution. Executing \rustinl{create_deque} symbolically, when we want to check if a write to \rustinl{Node.prev} field is safe, we do the same as what we did in concrete execution, except that instead of checking the existence of a $\predName{Node\_prev}$ chunk with a concrete value as the address we look for one with a term provably equal to $\sym{l}$ as its address. Both symbolic execution paths of \rustinl{create_deque} are safe. The safety of the path with successful allocation implies the safety of infinitely many corresponding concrete paths.

\subsection{Modular Symbolic Execution}
The preceding subsection showed how symbolic execution algorithm successfully verifies \rustinl{create_deque}. It also showed that after executing it there would be chunks of a \rustinl{Node} struct instance in the \heap{} at the address the function returns and the same address is stored in \rustinl{prev} and \rustinl{next} fields of that \rustinl{Node} instance in the heap. Moreover, the \rustinl{value} field is uninitialized. Now, what if we try to verify a program that calls \rustinl{create_deque} several times. Executing the body of functions over and over is a waste. Even worse, in the case of loops and recursive functions, our symbolic execution algorithm may not terminate. We also like to verify our programs in a modular way, \eg{} it is not pleasant to get involved with internal states of callees when we try to verify a caller. It would be useful, if we could save/document the knowledge we learn about the body of a function by symbolically executing it. Then instead of executing the body every time the function gets called, we can reuse that knowledge to infer what would be the state of execution if the call returns. This knowledge is called \emph{function contract}. Generally, we like a function's contract to tell us what is the weakest \emph{pre-condition}, \ie{} set of \emph{requirements}, for this function which if it holds no execution of the function exhibits \ubehav{}. That is, the minimal upper bound of the states if we execute the function's body starting from them, the execution would be safe. We also want the contract to tell us as much as possible about the effects that calling the function has on the execution state. In other words, what the strongest \emph{postcondition} the function \emph{ensures} is. That is, the maximal lower bound of guarantees about outcome states of all safe executions of the function. If a human/verifier provides us with a function contract in a well-defined logic, we can check the contract's propositions against the function body/implementation and if the body satisfies the contract, we can just reuse the contract every time we want to check a call to the function. This contract serves the same purpose as informal documentation, written in natural languages. But it is comprehensive and machine-checkable. Listing \ref{listing:create_deque_annot} shows \rustinl{create_deque} annotated with \verifast{} \sepLogic{} formulas as its contract.
\begin{listing}
\begin{minted}{Rust}
unsafe fn create_deque() -> *mut Node
//@ requires true;
/*@ ensures result!=0 &*& malloc_block_Node(result) &*& Node_prev(result, result) &*&
 Node_value(result, _) &*& Node_next(result, result);
 */
{
    let sentinel: *mut Node = std::alloc::alloc(std::alloc::Layout::new::<Node>()) as *mut Node;
    if sentinel.is_null() {
        std::alloc::handle_alloc_error(std::alloc::Layout::new::<Node>())
    }
    addr_of_mut!((*sentinel).prev).write(sentinel);
    addr_of_mut!((*sentinel).next).write(sentinel);
    return sentinel;
}
\end{minted}
\caption{\rustinl{create_deque} with contract, annotated in \verifast{} \sepLogic{}}
\label{listing:create_deque_annot}
\end{listing}

Let us verify an imaginary call to \rustinl{create_deque} with the contract shown in Listing \ref{listing:create_deque_annot}, using \emph{Modular Symbolic Execution}. First, we should verify that \rustinl{create_deque}'s body satisfies its contract. The \fContReq{} clause of the contract, \ie{} \rustinl{//@ requires true}, means to get executed safely, \rustinl{create_deque} needs that $true$ holds. Unsurprisingly, $true$ always holds in \sepLogic{}. So there are no special requirements, \ie{} no \heap{} chunks or facts about symbols, to assume when we start to verify the function. Also, \rustinl{create_deque} has no parameters, which means there is nothing in the $\sStore$ when we start checking its body. We start verifying \rustinl{create_deque}'s body from an empty $\sStore$, $\sHeap$, and $\sEPK{}$. In this specific case, we are starting from the same state as when we were executing just \rustinl{create_deque} symbolically and non-modularly. So the next three lines would have the same effect and we do not repeat those execution steps here. Although, there is an interesting difference at the return point. The contract's \fContEns{} clause, \ie{} \rustinl{//@ ensures result!=0 &*& malloc_block_Node(result) &*& ...}, is describing the effect of a call to \rustinl{create_deque} on the state of the caller, assuming the requirements of the call have been satisfied. So the return point is the point where we should verify the \fContEns{} clause. One of the facts this \fContEns{} clause asserts is that when a call to \rustinl{create_deque} returns, its mentioned chunks have been added to the \heap{}. The \rustinl{result} keyword in the \fContEns{} clause is a binder for the return value of the function, here, the symbolic value stored in \rustinl{sentinel}, \ie{} $\sym{l}$. To verify the \fContEns{} clause we \emph{consume} its mentioned chunks from the $\sHeap$. That is, we check the existence of the claimed chunks and since their access rights are being transferred to the caller, we deprive \rustinl{create_deque} of those rights by removing the chunks from $\sHeap$. It prevents us from transferring access rights of some \heap{} chunks to the caller twice. The \fContEns{} clause also mentions a persistent fact, \ie{} \rustinl{//@ ensures result!=0}, which we should check. The check is trivial because the exact assertion is in $\sEPK{}$ at the return point. In our example, after consuming the \fContEns{} clause chunks, $\sHeap$ would be empty. It means we could be sure that \rustinl{create_deque} does not leak memory chunks. The caller knows about the \fContEns{} clause chunks and the responsibility of deallocating them is now upon the higher-level code. \rust{}'s type system does not provide any guarantees about memory leaking in the presence of \kwunsafe{} code and tracking it is an added value of our \msexec{} algorithm.
Now we verified that the contract holds. Let us see what happens when we try to verify the call to \rustinl{create_deque} assuming the state at the call site is empty. By \rustinl{create_deque}'s contract, we know it does not need anything special before calling it. So we are good to go. We do not look up anything about \rustinl{create_deque}'s body. The next step of our \msexec{} algorithm is to just look up \rustinl{create_deque}'s contract and \emph{produce} the \fContEns{} clause. Assuming we represent the return value by $\sym{r}$, it leads to adding $\sym{r}\neq\rustinmath{0x0}$ to $\sEPK{}$ and adding the memory chunks $\mallocBlock{Node}(\sym{r}), \predName{Node\_prev}(\sym{r},\sym{r}), \predName{Node\_value}(\sym{r},\poison{}), \predName{Node\_next}(\sym{r},\sym{r})$ to the $\sHeap$. It captures the effect of the call to \rustinl{create_deque} and we can continue the execution of the rest of the caller's body.

\subsection{Modular Symbolic Execution and Verifying Safe Abstractions}
As we mentioned at the beginning of this section the \name{Deque} example is simple. That is because first, its interface is completely \kwunsafe{} and second, it interacts just using raw pointers. This simplicity of interface types helped us to establish the idea of \msexec{}. It also made us annotate the contract ourselves. In \rust{}, many facts about a function's contract are encoded in the function's type. In safe \rust{}, the type checker checks the safety of calls to the functions against the information encoded in their types, not an annotated contract. The type checker assumes the body of the function complies with its type. For purely safe functions this assumption gets checked during the type checking of the function itself. When it comes to safe abstractions, it is the programmer's responsibility to make sure that the function body complies with its type. Instead of verifying statically checked safe code, it is better to just verify that safe abstractions bodies satisfy the propositions encoded in their types. To verify a function's body, we start verifying the body from a symbolic state described by the function's contract \fContReq{} clause and check the validity of its contract's \fContEns{} clause at its return point(s). Now that the contract is encoded in the function's type, we need to represent the meaning of the \rust{}'s types in \sepLogic{} to use them in the \msexec{} algorithm.

To interpret the encoded information in a function type and use them in \msexec{}, we use the semantic model provided by \rustBelt{} \cite{Jung2017rustbelt}. In the next section, we explain \rustBelt{} briefly and using an example we represent our plan for Modular Symbolic Execution of safe abstractions based on \rustBelt{}'s semantic model for \rust{}'s types.

\section{RustBelt}\label{sec:rustbelt}
\rustBelt{} \cite{Jung2017rustbelt}, \rustHorn{} \cite{Matsushita2020rusthorn}, and \oxide{} \cite{Weiss2019oxide} are all well-known formal works around \rust{}. They all suggest calculi that capture \rust{}'s essence. However, we found \rustBelt{} more suitable for our purposes. \rustBelt{} proves \rust{}'s type safety taking \kwunsafe{} \rust{} into account, while the two other works do not. To prove the safety of \rust{} with \kwunsafe{} code, the popular \emph{Progress and Preservation} method is not useful. \kwunsafe{} \rust{} is not well-typed respecting safe \rust{} type system rules and \rust{} with relaxed typing rules for \kwunsafe{} code is not type-safe! That is why \rustBelt{} follows the semantic approach using \emph{logical relations} to prove the safety of \rust{} programs with \kwunsafe{} code. \rustBelt{} introduces \lambdaRust{}, a formal language close to \rust{}'s \emph{Mid-level Intermediate Representation} (\mir{}). Next, it provides a formal interpretation for \lambdaRust{}'s types and typing judgments in a dialect of \sepLogic{}, \iris{} \cite{IrisProject}. This interpretation is the semantic model they provide for \lambdaRust{}'s type system. Then they prove the safety of \lambdaRust{} using this semantic model following three steps, which have been mentioned in \rustBelt{} \cite{Jung2017rustbelt} paper as follows.
\begin{enumerate}
    \item \textquote{Verify that the typing rules of \lambdaRust{} are sound when interpreted semantically, \ie{} as lemmas establishing that the semantic interpretations of the premises imply the semantic interpretation of the conclusion. This is called the \emph{\name{fundamental theorem of logical relations}}.}
    \item \textquote{Verify that, if a closed program is semantically well-typed according to the model, its execution will not exhibit any unsafe/undefined behaviours. This is called \emph{\name{adequacy}}.}
    \item \textquote{For any library that employs \kwunsafe{} code internally, verify that its implementation satisfies the predicate associated with the semantic interpretation of its interface, thus establishing that the \kwunsafe{} code has indeed been safely \emph{encapsulated} by the library’s API. In essence, the semantic interpretation of the interface yields a library-specific verification condition.}
\end{enumerate}
With fundamental and adequacy theorems together, we have that \emph{syntactically well-typed programs are safe}. In comparison with the syntactic approach for safety proofs, \ie{} Progress and Preservation, there is an indirection in this semantic proof style. Intuitively, in progress and preservation, we show syntactically well-typed programs are safe, but here we show syntactically well-typed programs are semantically well-typed and then, semantically well-typed programs are safe. This indirection requires us to define a semantic model and makes the proof longer and harder. The reward of this extra effort, however, is that by the Adequacy theorem we can also show the safety of programs that are just semantically well-typed. This is the case mentioned in the third step of \rustBelt{}'s safety proof above.

Intuitively, in our approach using \msexec{}, we are following \rustBelt{}'s step three. By our \msexec{} we are proving no execution of functions of the \kwunsafe{} applying library violates their type's meaning. We will talk about the differences between our approach and \rustBelt{}, later in the Subsection \ref{ssec:added_value_rustbelt}. The semantic model \rustBelt{} provides is exactly what we needed in Section \ref{sec:mod_sym_exec} as the formal meaning of the interface of a safe abstraction.
To be precise, \iris{} which \rustBelt{} uses to represent its semantic model is not just a logic. It is a framework for higher-order concurrent separation logic that can be used for reasoning about the safety of concurrent programs. The fact that \rustBelt{} is also using \sepLogic{} for its semantic model, makes it easier for us to use. Recall that we are using a dialect of \sepLogic{} in our \msexec{} as well. In the next Subsection, we discuss using \rustBelt{}'s semantic model in our \msexec{} algorithm.

\subsection{RustBelt's semantic model and \msexec{}}\label{ssec:rustbelt_and_mse}
\begin{listing}
\begin{minted}{Rust}
    pub fn set<'a>(&'a self, n: i32) {
        let value_mut_ptr = &self.value as *const i32 as *mut i32;
        unsafe {
            *value_mut_ptr = n;
        }
    }
\end{minted}
\caption{A safe abstraction method}
\label{listing:cell_set}
\end{listing}

Listing \ref{listing:cell_set} shows the method \rustinl{set} of our simplified \rustinl{Cell} implementation shown in Listing \ref{listing:cell}. It has a lifetime parameter \rustinl{'a}, and two normal parameters. The interesting one is \rustinl{&'a self}. It is a shorthand for \rustinl{self: &'a Self} and \rustinl{Self} in our case is \rustinl{Cell}. Our de-sugared parameter would be \rustinl{self: &'a Cell}, a parameter named \rustinl{self} of type \rustinl{&'a Cell}, \ie{} a shared reference. A reference type carries much more information than a raw pointer. \rustinl{self}'s type tells us the following.
\begin{enumerate}
    \item Until the end of the time period denoted by lifetime \rustinl{'a}, the following guarantees hold:
    \item The parameter \rustinl{self} carries an aligned non-null address.
    \item There are enough bytes to store a \rustinl{Cell} value allocated at the address stored in \rustinl{self}.
    \item There is a valid \rustinl{Cell} value stored there.
    \item The memory region does not overlap with any memory region, owned by any active owning variable or referred to by any active mutable reference, \ie{} the memory would not get mutated by anyone. Although, other shared references to the memory region may exist, \eg{} other threads may read it.
\end{enumerate}
We need this information in a formal form. Let us go through \rustBelt{}'s semantics for this shared pointer briefly. In \rustBelt{} \textcquote{Jung2017rustbelt}{Each type $\tau$ is interpreted by a tuple $\rbInterpTuple{}$ of a natural number and two \iris{} predicates}. Listing \ref{listing:rustbelt_sem_shr_cell} shows \rustBelt{}'s predicates used for interpreting \rustinl{&'a Cell} type.
\begin{listing}
\begin{align}
&\rbInterpShrRefCellSize\label{eq:rustbelt_shr_ref_sz_prd}\\
&\rbInterpShrRefCellOwn\label{eq:rustbelt_shr_ref_own_prd}\\
&\rbInterpCellShr\label{eq:rustbelt_cell_shr_prd}
\end{align}
\caption{\rustBelt{}'s predicates related to interpreting a shared reference to \rustinl{Cell} type\protect\footnotemark}
\label{listing:rustbelt_sem_shr_cell}
\end{listing}
\footnotetext{Some details has been dropped for simplicity. For complete definitions see \cite{Jung2017rustbelt_technical}.}

Definition \ref{eq:rustbelt_shr_ref_sz_prd} of the $\rbPredSize{}$ value for shared references to $\tau$ under lifetime $\rbLft{\kappa}$ shows that all shared references are of size 1 memory unit. Definition \ref{eq:rustbelt_shr_ref_own_prd} of the $\rbPredOwn{}$ predicate for shared references to $\tau$ under lifetime $\rbLft{\kappa}$ has an interesting meaning. Its body uses the $\rbPredShr{}$ component of the interpretation of type $\tau$, \ie{} $\rbInterpTauShrPred$. This represents the fact that to have a shared reference to a type $\tau$ has different meanings depending on $\tau$. That is why \rustBelt{} defines a $\rbPredShr{}$ component for the interpretation of every type\footnote{We are not showing the definition of the component $\rbPredShr{}$ for shared references. It is not of interest in this example.}. Continuing to explore the meaning of predicate $\rbPredOwn{}$ for our shared reference to a \rustinl{Cell}, we need the definition of predicate $\rbPredShr{}$ of \rustinl{Cell}'s interpretation. It is shown in Definition \ref{eq:rustbelt_cell_shr_prd}. Before we explain it we need to know about \rustBelt{}'s \emph{\name{lifetime logic}}.

To facilitate expressing and reasoning about temporary and potentially shared ownership of resources in \iris{}, \rustBelt{} introduces a \name{lifetime logic} as an \iris{} library. To introduce these different kinds of ownership, this library relies on \emph{\name{borrow}}s, which are proposition constructors. The notation $\rbPredNaPersBr{\rbLft{\kappa}}{t}\dots$ is a kind of borrow named \emph{\name{non-atomic persistent borrow}} that represents thread-dependent temporary and potentially shared ownership. It is used to interpret the \rustinl{Cell} type. Let us explore the information this borrow and lifetime logic rules represent about \rustinl{Cell}. We need to know about them to explain the \msexec{} of \rustinl{Cell::set}.

Recall that the type \rustinl{Cell} allows clients to mutate its contents through a shared reference. That happens by applying an \kwunsafe{} superpower in its \rustinl{set} method. Having a shared reference does not rule out aliasing. So mutating data through shared references suggests the possibility of data races. To keep \rustinl{Cell} usages safe, we should make sure all of its aliases remain in the same thread. Fortunately, the type system takes care of it. The code line \rustinl{impl !Sync for Cell {}}, means values of type \rustinl{Cell} are not \rustinl{Sync}. That means they cannot be accessed simultaneously from different threads. In the \rust{} type system it means values of type \rustinl{&'a Cell} are not \rustinl{Send}, \ie{} shared references to values of type \rustinl{Cell} are not send-able to other threads. Moreover, no public function in \rustinl{Cell} leaks a deep reference to its contents. These facts together, prevent concurrent accesses to the memory owned by a \rustinl{Cell} and safe world can use \rustinl{Cell} without worrying about data races.

In \rustBelt{} a type $\tau$ is \rustinl{Send}, if and only if, the $\rbInterp{\tau}.\rbPredOwn{}(t,\overline{\upsilon})$ definition does not depend on the thread identifier $t$. A type $\tau$ is \rustinl{Sync}, if and only if, the type of shared references to $\tau$, \ie{} $\rbShrRef{\rbLft{\kappa}}\ \tau$, is \rustinl{Send}. The fact that \rustinl{Cell} is not \rustinl{Sync} has been reflected in \rustBelt{}'s interpretation as follows. The $\rbPredNaPersBr{\rbLft{\kappa}}{t}$ which has been used in the $\rbPredShr{}$ component of $\rbInterp{\mathbf{cell}}$ depends on the thread identifier $t$. In short \rustinl{Cell}'s sharing predicate depends on the thread identifier. Since $\rbInterp{\rbShrRef{\rbLft{\kappa}} \tau}.\rbPredOwn{}$, shown in the Definition \ref{eq:rustbelt_shr_ref_own_prd}, consists of $\rbInterp{\tau}.\rbPredShr{}$, $\rbInterp{\rbShrRef{\rbLft{\kappa}} \mathbf{cell}}.\rbPredOwn{}$ depends on $t$ as well, reflecting that shared references to \rustinl{Cell} are not \rustinl{Send}.

The interesting point in proving \rustBelt{}'s step three about \rustinl{Cell::set} is that we need full/write access to \rustinl{Cell}'s content to be sure the write operation is safe. To understand how we can obtain such access, we need to look at the \name{lifetime logic}'s rules that provide us access to the resources held by a borrow. In our example, the resources held by a non-atomic persistent borrow. Listing \ref{listing:rustbelt_lftl_na_acc} shows rule \ruleName{LftL-na-acc} of lifetime logic. This is the rule we are looking for.
\begin{listing}
\begin{align}
\rbRuleLftlNaAcc\label{eq:rustbelt_lftl_na_acc}
\end{align}
\caption{\ruleName{LftL-na-acc} rule from \rustBelt{}'s lifetime logic}
\label{listing:rustbelt_lftl_na_acc}
\end{listing}

It describes how we can get full access to a resource $\prop$ when we have it under a non-atomic persistent borrow. Besides $\rbRuleLftlNaAccBor{}$ itself, the rule requires $\rbRuleLftlNaAccLftTk{}$ and $\rbRuleLftlNaAccNaTk{}$. Intuitively, in the \rustinl{Cell::set} example if we provide a witness that lifetime \rustinl{'a} is alive and we are in the same thread that the \rustinl{Cell} itself is we can get our full access. But there is more than that about $\rbRuleLftlNaAccLftTk{}$ and $\rbRuleLftlNaAccNaTk{}$. Let us explain them in order.

$\rbLftTokDef{}$ is the lifetime logic's \emph{\name{lifetime token}}, representing lifetime $\rbLft{\kappa}$ is alive/ongoing. That is the same lifetime as the one that appears in the non-atomic persistent borrow itself. To give us the resource $\prop$, this rule requires us to provide evidence that the borrow lifetime is alive; fair enough. The fraction $q$, such that $0<q\leq1$, in the lifetime token plays an important role. Whenever a lifetime starts, we get its token with the full fraction, $\rbLftTok{\rbLft{\kappa}}{1}$. The lifetime logic's rules about accessing \name{borrow}s consume a fraction of the lifetime token for a borrow's lifetime, besides other requirements, to provide us with:
\begin{enumerate}
    \item Access to the resources behind the borrow. Represented in \ruleName{LftL-na-acc} by $\prop$.
    \item An \emph{\name{update}} which takes back the borrowed resource and gives back the lifetime token fraction that had been used when the rule was applied to provide the resource. In the case of \ruleName{LftL-na-acc} the $\rbRuleLftlNaAccUpdate{}$ part.
\end{enumerate}
In lifetime logic, we cannot show a lifetime $\rbLft{\kappa}$ is ended unless we consume its token with the full fraction. It means we need to take back all the fractions that have been used to get access to resources behind borrows under $\rbLft{\kappa}$. Taking the fractions back is just possible through those updates we just mentioned, in the case of \ruleName{LftL-na-acc} the $\rbRuleLftlNaAccUpdate{}$. Those updates always need the resources they have handed out, back. That is, to end a lifetime, we are forced to make sure all the permissions granted through borrows under that lifetime have been taken back. Intuitively, the aliveness of a lifetime is a credit, we borrow access to resources relying on that lifetime and to end that lifetime we should have paid our debts to the lifetime back.

Moreover, the rule requires the \name{non-atomic token} $\rbNonAtomTok{t}$, bound to the same thread as the non-atomic persistent borrow. \textcquote{Jung2017rustbelt}{This token is created at the birth of the thread, and threaded through all of its control flow. That is, every function receives it and has to return it.} The same scenario of consumption and giving back of $\rbLftTokDef{}$ in \ruleName{LftL-na-acc} happens for $\rbNonAtomTok{t}$ too. It means at return points we need $\rbNonAtomTok{t}$ back and to have that again we need to give back the resource we have granted using \ruleName{LftL-na-acc} relying on the fact that we are in thread $t$. Intuitively, at the function's return point, it gets checked that whatever thread-dependent resource has been taken, has been given back.

Back to our \msexec{} algorithm, starting from a symbolic state containing \rustBelt{}'s predicates extracted from \rustinl{Cell::set}'s type, we should be able to extract the facts we need to verify \rustinl{Cell::set}'s body. Moreover we need to check the integrity of the type system invariant at return points. To keep the text concise, we skip the details. Using what we learned from \rustBelt{}'s semantic model and its lifetime logic, the outline of our \msexec{} for safe abstraction \rustinl{Cell::set} would be as follows:
Since, by \rust{}'s type system, it is always guaranteed that the instantiations of a function's lifetime parameters outlive the function execution period, at the beginning of the function, we have a fraction of the lifetime token for each lifetime parameter. The function's execution period is a lifetime, always shown by binder $\rbFnLft{}$. Obviously, function execution is happening in a thread; so we get a non-atomic token for the current thread. And of course, we get the $\rbPredOwn{}$ component of the interpretation of the type of the function's parameters. That gives us the symbolic execution state, shown in row number \ref{tabrow:msexec_cell_set_start} of Table \ref{tab:msexec_cell_set}, to start our symbolic execution\footnote{To show our purpose clearer, we dropped details regarding the facts that in \rustBelt{} there is no mutable store and all locals, \ie{} parameters and local variables, are owned pointers. We are just showing them here as store variables.}.
\begin{table}[H]
\centering
\caption{Modular Symbolic Execution of the safe abstraction method \rustinl{Cell::set}.\\For all rows $\sStore{}=\{\symSVar{self}:\sym{s}, \symSVar{n}:\sym{n}\}$ and $\sEPK{}=\{\rbFnLft{}\rbLftInc{}\sym{\rbLft{a}}, 0<\sym{q}\le1\}$.}\label{tab:msexec_cell_set}
\begin{tabular}{|N|c|l|}
\hline
\multicolumn{1}{|c|}{\#} & \rust{} & $\sRsc{}$ \\ \hline
\label{tabrow:msexec_cell_set_start}
&\rustinl{fn set<'a>(...)}
&$\rbNonAtomTok{\sym{t}}, \rbLftTok{\sym{\rbLft{a}}}{\sym{q}}, \rbPredInterpShrRefCellOwnSymb$ \\ \hline
\label{tabrow:msexec_cell_set_open1}
&\rustinl{//@open shr.own}
&$\rbNonAtomTok{\sym{t}}, \rbLftTok{\sym{\rbLft{a}}}{\sym{q}}, \rbPredInterpCellShrSym$ \\ \hline
\label{tabrow:msexec_cell_set_open2}
&\rustinl{//@open cell.shr}
&$\rbNonAtomTok{\sym{t}}, \rbLftTok{\sym{\rbLft{a}}}{\sym{q}}, \rbCellShrSymb$ \\ \hline
\label{tabrow:msexec_cell_set_lftl_na_acc}
&\rustinl{//@lemma lftl_na_acc}
&\makecell[l]{$\rbRuleLftlNaAccOutcomeRscSym,$ \\ $\rbRuleLftlNaAccOutcomeUpdateSym$} \\ \hline
\label{tabrow:msexec_cell_set_write}
&\rustinl{*value_mut_ptr = n;}
&\makecell[l]{$\rbRuleLftlNaAccOutcomeRscSymWritten,$ \\ $\rbRuleLftlNaAccOutcomeUpdateSym$} \\ \hline
\label{tabrow:msexec_cell_set_update}
&\rustinl{//@apply update s|->n}
&$\rbNonAtomTok{\sym{t}}, \rbLftTok{\sym{\rbLft{a}}}{\sym{q}}$ \\ \hline
\end{tabular}
\end{table}

To justify the write in \rustinl{Cell::set} we need write permission for the \rustinl{Cell}'s content. We can get access to corresponding memory chunks by opening the $\rbPredInterpShrRefCellOwnSymb$ to its definition which gives us $\rbPredInterpCellShrSym{}$. By opening the latter again, we would have the symbolic execution state in the row number \ref{tabrow:msexec_cell_set_open2} in Table \ref{tab:msexec_cell_set}.

Now using \ruleName{LftL-na-acc} shown in Listing \ref{listing:rustbelt_lftl_na_acc} we can get write access. But recall that the rule also needs to consume a fraction of borrow lifetime token, \ie{} $\rbLftTok{\sym{\rbLft{a}}}{\sym{q'}}$, and the non-atomic token bound to the current thread, \ie{} $\rbNonAtomTok{\sym{t}}$. Because we do not need $\rbLftTok{\sym{\rbLft{a}}}{}$ for the rest of \rustinl{Cell::set} body to get access to another borrow, we can just give all the fraction of $\rbLftTok{\sym{\rbLft{a}}}{}$ we have to \ruleName{LftL-na-acc}. After applying the rule we have the symbolic state shown in the row number \ref{tabrow:msexec_cell_set_lftl_na_acc} in Table \ref{tab:msexec_cell_set}.

The write can be verified now because we have full access to the \heap{} chunk $\sym{s}\mapsto\overline{\upsilon}$. The write operation updates the value of the chunk giving us the updated resource $\rbRuleLftlNaAccOutcomeRscSymWritten$. The state is shown in the row number \ref{tabrow:msexec_cell_set_write} of Table \ref{tab:msexec_cell_set}. By the next statement, \rustinl{Cell::set} returns. \rustinl{Cell::set}'s return type is not shown explicitly which in \rust{} means it is \rustinl{()}, \ie{} the unit type. To close $\rbInterp{()}.\rbPredOwn{}(\sym{t}, [])$ does not need any resources so we can easily close it out of thin air. There is no destructor call happening here as well.
As a check for preserving the type system invariant at the return point, we consume whatever fraction of external lifetime tokens we got for lifetime parameters. In the case of \rustinl{Cell::set} there is just \rustinl{'a}. So we need to consume back $\rbLftTok{\sym{\rbLft{a}}}{\sym{q}}$. By doing so we make sure whatever resources we have granted from borrows under \rustinl{'a}, we are giving back to the caller. Recall that to have $\rbLftTok{\sym{\rbLft{a}}}{\sym{q}}$ and $\rbNonAtomTok{\sym{t}}$ back, we need to use the update $\rbRuleLftlNaAccOutcomeUpdateSym{}$ in our $\sRsc{}$. Using the update needs consuming the granted resource $\rbRuleLftlNaAccOutcomeRscSym{}$, \ie{} giving it back. The caller needs to take back the lifetime token fraction provided to call the current function. Another obvious return point verification is consuming the non-atomic token with the current thread binder, $\rbNonAtomTok{\sym{t}}$. Recall it is being threaded through all the calls in a thread.

Our target claim is that, for a \emph{type-checked} program, if the \msexec{} algorithm successfully executes all safe abstractions and the whole \kwunsafe{} hierarchy of code behind them, no execution of that program will exhibit \ubehav{}. In \rustBelt{}'s terminology, that means if our \msexec{} algorithm verified a safe abstraction, there exists a \rustBelt{} proof to show the safe abstraction holds its interface type guarantees. In short, we intend for our \msexec{} algorithm to be sound regarding to step three of \rustBelt{}'s safety proof mentioned at the beginning of this section.

\section{Implementation}\label{sec:impl}
To evaluate our \msexec{} algorithm on non-trivial examples and case studies, we are implementing our algorithm to have a tool to symbolically execute \rust{} programs. There are two important questions needed to be addressed regarding our implementation. First, which representation of \rust{} we should symbolically execute and second, how we can reuse the capabilities of the existing research tool \verifast{} to implement our algorithm.

\subsection{Executing \mir{}}
Surface \rust{} has a heavily sugared syntax and there is no formal operational semantics by the language community for it. \mir{}, however, is heavily simplified by the compiler. In \mir{}, temporary values of higher representations of \rust{} programs are bounded and function bodies are represented in the form of a \name{Control-flow Graph}. But the essence of ownership and borrowing representing types is still preserved in this intermediate representation. \name{Generic} definitions are also still in place in \mir{}. Therefore, it is much simpler and easier to execute and reason about \mir{} instead of surface \rust{} while having interesting properties of language in hand to work with. Both \rustBelt{} and \rustHorn{} calculi, \lambdaRust{} and \corLang{} respectively, are inspired by \mir{} witnessing this fact. Moreover, to compensate for the lack of formal operational semantics, the language community relies on a \mir{} interpreter named \miri{}. It is much easier to refer to \miri{} to see what exactly the semantics of a program is. That is why we decided to symbolically execute \mir{} representation in the background.
To get the \mir{} representation of a program along with type definitions and user annotations, we have implemented a \rust{} program which uses the official \rust{} compiler front-end to type and borrow check the program and generate its \mir{}. Using the official compiler front-end saves a lot of work and also prevents our tool to diverge from what exactly the \rust{} compiler is. If the program passes the front-end checks successfully, our tool translates all required information to \capnp{} \cite{capnp} data structures and dumps it to standard output. \capnp{} is a data interchange format supported in many different programming languages. This makes our \mir{} extraction program reusable for other \rust{} analyser tools.

\subsection{Executing \mir{} in \verifast{}}
Fortunately, we do not need to implement a symbolic execution tool capable of reasoning about \sepLogic{} propositions from scratch. \verifast{} is a research tool for verifying \C{} and \java{} programs annotated with \verifast{}'s dialect of \sepLogic{} and \verifast{}'s ghost commands. Extending \verifast{} to support \rust{}, or more accurately to support \mir{}, spares us implementing the executing and reasoning engine from scratch. To symbolically execute \mir{} in \verifast{}, our approach is to translate \mir{}, \rust{}'s types semantics, and user annotations together into \verifast{}'s \C{} abstract syntax tree (\astree{}). By doing so, we are effectively defining an operational semantics for \mir{} using \verifast{}'s \C{} operational semantics. A similar process of defining operational semantics for \lambdaRust{} by translating it to another language happens in \rustBelt{}. \textcquote{Jung2017rustbelt}{The operational semantics of \lambdaRust{} is given by translation into a core language. The core language is a lambda calculus equipped with primitive values, pointer arithmetic, and concurrency}.

Since \mir{} is a control-flow graph, translating the code control-flow to \C{} control constructs is straightforward. For some data types, there are direct equivalents, \eg{} \rustinl{bool} and more or less integers; some others do not have direct equivalents but it is still easy to translate them. As an example, the approach for translating \name{tuples} is using \C{} \mintinline{C}{struct}s with reserved names. For more complex \rust{} types that are not fully representable by \C{} types, as already mentioned, the approach is to add \rustBelt{} type semantics represented in \verifast{}'s \sepLogic{}. The examples in appendix \ref{sec:appx_intended_verifast} illustrate our intention for generating \rustBelt{} rules and predicates for a safe abstraction\footnote{The mentioned examples have been provided by Prof. Bart Jacobs.}.

\begin{figure}
    \centering
    \includegraphics[width=0.75\textwidth]{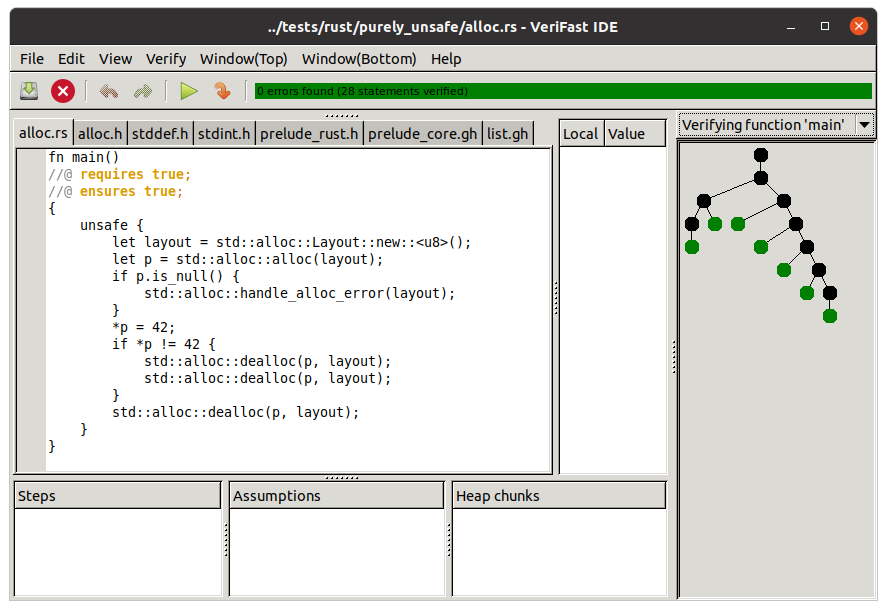}
    \caption{The alloc.rs \rust{} program verified by \verifast{}}
    \label{fig:verifast_alloc_example}
\end{figure}

At the time of writing this report, the tool can verify a simple example of memory allocation, access and un-allocation, shown in Figure \ref{fig:verifast_alloc_example}. Even this simple example includes two \name{generic} functions whose definitions are parameterised by a type. The instantiations of functions \rustinl{new} and \rustinl{is_null} used in the example are \rustinl{std::alloc::Layout::new::<u8>()} and \rustinl{std::ptr::mut_ptr::<impl *mut u8>::is_null(*mut u8)} respectively. Generic definitions are not generally handled yet. For these cases, we substitute with equivalents of their instantiated implementation.

The \mir{} extraction program and the \verifast{} extension for supporting \rust{} are works in progress and currently support a very limited subset of \rust{}. The development of \verifast{} including the \mir{} extractor program is being done in branch \textsf{\name{rust}} in a fork of \verifast{} that can be found at \url{https://github.com/Nima-Rahimi-Foroushaani/verifast}. The current status of the code including the \textsf{\name{alloc}} example shown in Figure \ref{fig:verifast_alloc_example} is available as a \name{Zenodo} drop at \url{https://doi.org/10.5281/zenodo.7472607}. To build and run the code follow the instructions provided along with the Zenodo drop.

\subsection{Added value with respect to RustBelt}\label{ssec:added_value_rustbelt}
A valid question then is that while \rustBelt{} already exists why should we bother to enhance \verifast{} to verify \rust{} programs with \kwunsafe{} code. To verify the safety of a new library with \rustBelt{} one would need to have considerable knowledge about \iris{} in the first place. Moreover, it would be necessary to translate the surface \rust{} code to \lambdaRust{}. After all, it is just the starting point to the safety proof of the program. In our approach, however, the required knowledge is \verifast{} separation logic and our intended encoding of the \rustBelt{} semantic framework including lifetime logic in \verifast{}. \verifast{} would work with the surface \rust{} and the translation to \mir{} happens in the background using the \rust{} compiler front-end. That reduces the burden of learning for \rust{} developers who aim to verify their code.
On the other hand, our approach leads to having actual \rust{} code and \verifast{} annotation, \ie{} verifiable formal documentation, together in the same place. Our hypothesis is that it leads to a better information encoding scheme for practicality. Listing \ref{listing:annotated_unsafe_swap} shows an actual \kwunsafe{} function from the \rust{} core library with a hypothetical \verifast{} annotation along with a part of corresponding informal documentation.
\begin{listing}
\begin{minted}[breaklines]{Rust}
/// ...
/// Behavior is undefined if any of the following conditions are violated:
/// * Both `x` and `y` must be [valid] for both reads and writes of `count *
///   size_of::<T>()` bytes.
/// * Both `x` and `y` must be properly aligned.
/// * The region of memory beginning at `x` with a size of `count *
///   size_of::<T>()` bytes must *not* overlap with the region of memory
///   beginning at `y` with the same size.
/// ...
pub const unsafe fn swap_nonoverlapping<T>(x: *mut T, y: *mut T, count: usize)
//@ requires Interp_own(T)(x,?vs1) &*& Interp_own(T)(y,?vs2) &*& length(vs1)==count &*& length(vs2)==count
//@ ensures Interp_own(T)(x,?vs2) &*& Interp_own(T)(y,?vs1) &*& length(vs1)==count &*& length(vs2)==count
{...}
\end{minted}
\caption{An \kwunsafe{} function from \rust{} core library with a hypothetical \verifast{} annotation}
\label{listing:annotated_unsafe_swap}
\end{listing}

\section{Future Plans}\label{sec:future}
In subsection \ref{ssec:added_value_rustbelt}, we mentioned some practical added value for verifying \kwunsafe{} \rust{} using \verifast{} in comparison with \rustBelt{}. But we plan to contribute further to the safety of \rust{} ecosystem in other ways as well in the future. In subsection \ref{ssec:rig_soundness} we explain the possibilities of further formal work to establish the soundness of our \msexec{} algorithm.
One of the problems we are targeting to address in \verifast{} is the safety problems that occur in the presence of \kwunsafe{} code and stack unwinding. In subsection \ref{ssec:panic_safety} we discuss the problem and why our implementation shows promise to solve that.

\subsection{Rigorous Soundness}\label{ssec:rig_soundness}
One could rightfully argue about the soundness of our \msexec{} algorithm respecting \rustBelt{} proofs. To support our soundness claim rigorously, there are two possible approaches. One is to formalize our \msexec{} algorithm based on \lambdaRust{}'s operational semantics and prove that if it verifies a function there is a \rustBelt{} proof for the safety of the function as well. Another approach is to generate a function-specific \iris{} proof out of executing the function. For that, we need to define a function between a passed/verified symbolic execution tree of a function and a \rustBelt{} soundness proof about it.

\subsection{Panic Safety and Stack Unwinding}\label{ssec:panic_safety}
According to \name{The Rustonomicon} \cite{rustonomicon}, \rust{}'s error handling scheme is as follows:
\begin{itemize}
    \item If something might reasonably be absent, \rustinl{Option} is used.
    \item If something goes wrong and can reasonably be handled, \rustinl{Result} is used.
    \item If something goes wrong and cannot reasonably be handled, the thread panics.
    \item If something catastrophic happens, the program aborts.
\end{itemize}
Although, the first two, are recommended and common ways of reporting unhappy results, there are many places \rust{} code may panic. \textcquote{rustonomicon}{Panics cause the thread to halt normal execution and unwind its stack, calling destructors as if every function instantly returned}. A program can recover from panic and handle it using \href{https://doc.rust-lang.org/std/panic/fn.catch_unwind.html}{\rustinl{std::panic::catch_unwind}}. On the other hand, \href{https://doc.rust-lang.org/std/process/fn.abort.html}{\rustinl{std::process::abort}}, immediately terminates the current process. In the case of panic, the compiler takes care of the safety and the cleaning up in the unwinding execution path. Once again, when it comes to \kwunsafe{} code, the information encoded in types is not enough to be sure about safety. In presence of the \kwunsafe{} blocks, \textcquote{rustonomicon}{code that transiently creates unsound states must be careful that a panic does not cause that state to be used}. Listing \ref{listing:panic_safety} shows an example of such bugs, inspired by a real-life one \cite{RustIssues2015bheap_not_excp_safe}. This kind of bug is hard for a human to track. Programmers need to constantly keep the probability of panic in mind and address all of the transient unsound states.
\begin{listing}
\begin{minted}{Rust}
use core::mem::{replace, MaybeUninit};
use core::ptr;

pub struct BinaryHeap<T> {
    pub data: Vec<T>,
}

impl<T: Ord> BinaryHeap<T> {
    // T implements Ord

    pub fn sift_up(&mut self, start: usize, mut pos: usize) {
        unsafe {
            let new = replace(
                &mut self.data[pos],
                MaybeUninit::<T>::zeroed().assume_init(),
            );
            // There is an element with all bytes zeroed
            // which is not necessarily a valid value
            while pos > start {
                let parent = (pos - 1) >> 1;
                if new <= self.data[parent] {
                    // What if the '<=' panics!
                    break;
                }
                let x = replace(
                    &mut self.data[parent],
                    MaybeUninit::<T>::zeroed().assume_init(),
                );
                ptr::write(&mut self.data[pos], x);
                pos = parent;
            }
            ptr::write(&mut self.data[pos], new);
        }
    }
}
\end{minted}
\caption{An example of memory safety bug in presence of \kwunsafe{} code and function call panic inspired from \rust{}'s issue 25842 \cite{RustIssues2015bheap_not_excp_safe}}
\label{listing:panic_safety}
\end{listing}
Fortunately, the bug from the standard library has been fixed. But notice that it is a mistake made by experts. This kind of bug is still showing up now and then in the ecosystem. That is why \rudra{} \cite{Bae2021rudra} aims for this bug's pattern as one of its targets. While \rudra{} is a valuable static analyzer which has made the language ecosystem safer, it does not guarantee panic safety. The panic execution path becomes explicit once the compiler reduces surface \rust{} to \mir{}. Listing \ref{listing:panic_safety_mir} shows a part of the compiled down \mir{} for \rustinl{sift_up} that has been shown in Listing \ref{listing:panic_safety}. It shows \emph{Basic Block} \rustinl{bb8} where the call to function \rustinl{le}, \ie{} operator $\le$ gets executed. One of the possible successors of the \emph{Terminator} for this function call corresponds to the case if the function call panics and it is basically a jump to \emph{Basic Block} \rustinl{bb23}.

To address the panic safety in presence of \kwunsafe{} code, there are two possible steps to take. First we can extend \rustBelt{} with panics and prove the safety of safe abstractions in presence of panic there. Second, since in our tool we are symbolically executing \mir{} in the background, it can naturally take the panic execution paths into account. However, the unwinding path does not return a value from the function we are verifying. Then not all the guarantees the function type asserts, need to hold. We need to study what the exact necessary checks are to claim the \emph{exception safety} of a function after a panic.
\begin{listing}
\begin{minted}{Rust}
bb8: {
        _21 = _22;
        _19 = <T as PartialOrd>::le(move _20, move _21) -> [return: bb9, unwind: bb23];
    }
\end{minted}
\caption{Part of \mir{} corresponding to method \rustinl{sift_up} has shown in Listing \ref{listing:panic_safety}. Stack Unwinding execution path is explicit in \mir{}}
\label{listing:panic_safety_mir}
\end{listing}

\section{Conclusion}\label{sec:conc}
The problem of verifying the memory safety of \rust{} programs with \kwunsafe{} blocks suggests a good opportunity to contribute to the safety of the software industry. Our modular symbolic execution approach is inspired by the formal work \fwVerifast{} \cite{Jacobs2015fw_verifast}, relying on the semantic model provided by \rustBelt{} \cite{Jung2017rustbelt}. The solid formal foundation we are building upon makes our approach very likely to have solid results. On the other hand, in our research path, we keep evaluating our algorithm with real-life scenarios by extending \verifast{} and using \rust{} compiler front-end. \verifast{} as a verification software has proven to be useful. There is a fundamental interest in safety in the \rust{} community. Integrating the official \rust{} compiler with \verifast{} provides the possibility for \rust{} ecosystem to improve the safety of language.

\printbibliography[heading=bibintoc, title={bibliography}]

@article{Jung2017rustbelt,
    author = {Jung, Ralf and Jourdan, Jacques-Henri and Krebbers, Robbert and Dreyer, Derek},
    title = {RustBelt: Securing the Foundations of the Rust Programming Language},
    year = {2017},
    issue_date = {January 2018},
    publisher = {Association for Computing Machinery},
    address = {New York, NY, USA},
    volume = {2},
    number = {POPL},
    url = {https://doi.org/10.1145/3158154},
    doi = {10.1145/3158154},
    journal = {Proc. ACM Program. Lang.},
    month = {12},
    articleno = {66},
    numpages = {34}
}

@article{Jung2017rustbelt_technical,
    author = {Jung, Ralf and Jourdan, Jacques-Henri and Krebbers, Robbert and Dreyer, Derek},
    title = {RustBelt: Securing the Foundations of the Rust Programming Language – Technical appendix and Coq development},
    year = {2017},
    url = {https://plv.mpi-sws.org/rustbelt/popl18/}
}

@incollection{Matsushita2020rusthorn,
	doi = {10.1007/978-3-030-44914-8_18},
	url = {https://doi.org/10.1007%2F978-3-030-44914-8_18},
	year = 2020,
	publisher = {Springer International Publishing},
	pages = {484--514},
	author = {Yusuke Matsushita and Takeshi Tsukada and Naoki Kobayashi},
	title = {{RustHorn}: {CHC}-Based Verification for Rust Programs},
	booktitle = {Programming Languages and Systems}
}

@online{Jung2017mutexguard_unsound,
    author = "Ralf Jung",
    title = "\rustinl{MutexGuard<Cell<i32>>} must not be \rustinl{Sync}. Rust issue \#41622",
    url = "https://github.com/rust-lang/rust/issues/41622"
}

@inproceedings{Bae2021rudra,
    author = {Bae, Yechan and Kim, Youngsuk and Askar, Ammar and Lim, Jungwon and Kim, Taesoo},
    title = {Rudra: Finding Memory Safety Bugs in Rust at the Ecosystem Scale},
    year = {2021},
    isbn = {9781450387095},
    publisher = {Association for Computing Machinery},
    address = {New York, NY, USA},
    url = {https://doi.org/10.1145/3477132.3483570},
    doi = {10.1145/3477132.3483570},
    booktitle = {Proceedings of the ACM SIGOPS 28th Symposium on Operating Systems Principles},
    pages = {84–99},
    numpages = {16},
    location = {Virtual Event, Germany},
    series = {SOSP '21}
}

@online{thebook,
    author = "Steve Klabnik and Carol Nichols with contributions from the Rust Community",
    title = "The Rust Programming Language",
    url = "https://doc.rust-lang.org/book/title-page.html"
}

@online{rustonomicon,
    author = "Contributions from the Rust Community",
    title = "The Rustonomicon",
    url = "https://doc.rust-lang.org/nomicon"
}

@misc{Weiss2019oxide,
    author = {Weiss, Aaron and Gierczak, Olek and Patterson, Daniel and Ahmed, Amal},
    doi = {10.48550/ARXIV.1903.00982},
    url = {https://arxiv.org/abs/1903.00982},
    title = {Oxide: The Essence of Rust},
    publisher = {arXiv},
    year = {2019},
    copyright = {arXiv.org perpetual, non-exclusive license}
}

@online{RustIssues2015bheap_not_excp_safe,
    title = "\rustinl{BinaryHeap} is not exception safe. \rust{} issue \#25842",
    url = "https://github.com/rust-lang/rust/issues/25842"
}

@article{Jacobs2015fw_verifast,
    doi = {10.2168/lmcs-11(3:19)2015},
    url = {https://doi.org/10.2168%2Flmcs-11%283%3A19%292015},
    year = 2015,
%    month = {sep},
    publisher = {Centre pour la Communication Scientifique Directe ({CCSD})},
    volume = {11},
    number = {3},
    author = {Bart Jacobs and Fr{\'{e}}d{\'{e}}ric Vogels and Frank Piessens},
    editor = {Tobias Nipkow},
    title = {Featherweight {VeriFast}},
    journal = {Logical Methods in Computer Science}
}

@online{IrisProject,
    title = "\iris{}",
    url = "https://iris-project.org/"
}

@online{capnp,
    title = "\capnp{}",
    url = "https://capnproto.org/"
}

@online{verifast,
    title = "\verifast{}",
    url = "https://github.com/verifast/verifast"
}

\appendix
\section{Intended encoding of the \rustBelt{}'s semantic model in \verifast{}}{\label{sec:appx_intended_verifast}}
\textcolor{red}{The examples that have been discussed in this appendix, have been provided by Prof. Bart Jacobs, not by Nima Rahimi Foroushaani}\\
The example that has been shown in Listing \ref{listing:appx_cell_verifast_rust} is an illustration of our goal for verifying \rust{}'s safe abstractions using \verifast{}. The other example in Listing \ref{listing:appx_cell_verifast_c} shows the outcome of our intended translation from the example of Listing \ref{listing:appx_cell_verifast_rust} to a \C{} program plus required \rustBelt{}'s semantic model rules and predicates.

\begin{listing}
\begin{minted}{Rust}
pub struct Cell_i32 {
    value: i32
}

/*@

pred Cell_i32_nonatomic_borrow_content(l: *i32, t: thread_id)() =
  *l |-> _;

interp Cell_i32 {
  pred shared(k: lifetime, t: thread_id, l: *i32) = nonatomic_borrow(k, t, l, Cell_i32_nonatomic_borrow_content(l, t));
}

@*/

impl Cell_i32 {
    fn replace(&self, val: i32) -> i32
    //@ req [?q]lifetime(?a) &*& Cell_i32_shared(a, ?t, self) &*& thread_token(t);
    //@ ens [q]lifetime(a) &*& thread_token(t);
    {
        //@ open Cell_i32_shared(a, t, self);
        //@ open_nonatomic_borrow(a, t, self, q);
        //@ open Cell_i32_nonatomic_borrow_content(self, t)();
        let result: i32 = self.value;
        self.value = val; // using unsafe superpower
        //@ close Cell_i32_nonatomic_borrow_content(self, t)();
        //@ close_nonatomic_borrow();
        return result;
    }
}
\end{minted}
\caption{A \rustinl{Cell} implementation in \rust{} with the intended user provided \verifast{}'s annotations that are required for verifying it. This example has been provided by Prof. Bart Jacobs}
\label{listing:appx_cell_verifast_rust}
\end{listing}

\begin{listing}
\begin{minted}{Rust}
/*@

// Lifetime logic

abstract_type lifetime; // Type of lifetimes
abstract_type thread_id; // Type of thread IDs

predicate lifetime(lifetime k;); // Lifetime token
predicate thread_token(thread_id t); // nonatomic token with Top mask ([NaInv: t.Top] in RustBelt)

predicate nonatomic_borrow(lifetime k, thread_id t, void *l, predicate() P); // nonatomic borrow with mask Nshr.l

lemma void open_nonatomic_borrow(lifetime k, thread_id t, void *l, real q); // Rule LftL-na-acc with N = Nshr.l and requiring NaInv: t.Top instead of NaInv: t.N
    requires nonatomic_borrow(k, t, l, ?P) &*& [q]lifetime(k) &*& thread_token(t);
    ensures P() &*& close_nonatomic_borrow_token(P, q, k, t);

predicate close_nonatomic_borrow_token(predicate() P, real q, lifetime k, thread_id t);

lemma void close_nonatomic_borrow();
    requires close_nonatomic_borrow_token(?P, ?q, ?k, ?t) &*& P();
    ensures [q]lifetime(k) &*& thread_token(t);

// Cell<i32> type interpretation

predicate_ctor Cell_i32_nonatomic_borrow_content(void *l, thread_id t)() =
    integer(l, _);
predicate Cell_i32_shared(lifetime k, thread_id t, void *l) = // SHR predicate for Cell<i32>
    nonatomic_borrow(k, t, l, Cell_i32_nonatomic_borrow_content(l, t));

@*/

// fn replace<'a>(self: &'a Cell<i32>, val: i32) -> i32
int replace(int *self, int val)
//@ requires [?q]lifetime(?a) &*& Cell_i32_shared(a, ?t, self) &*& thread_token(t);
//@ ensures [q]lifetime(a) &*& thread_token(t);
{
  //@ open Cell_i32_shared(a, t, self);
  //@ open_nonatomic_borrow(a, t, self, q);
  //@ open Cell_i32_nonatomic_borrow_content(self, t)();
  int result = *self;
  *self = val;
  //@ close Cell_i32_nonatomic_borrow_content(self, t)();
  //@ close_nonatomic_borrow();
  return result;
}
\end{minted}
\caption{The intended \C{} translation of the example, shown in Listing \ref{listing:appx_cell_verifast_rust} with the \verifast{}'s annotations. The annotations here are the user provided ones in the example shown in Listing \ref{listing:appx_cell_verifast_rust} plus the ones that our intended approach would generate. This example has been provided by Prof. Bart Jacobs}
\label{listing:appx_cell_verifast_c}
\end{listing}

\end{document}